\documentclass[conference, table]{IEEEtran}
\IEEEoverridecommandlockouts

\usepackage{cite} % needed to identify funding in the first footnote
\usepackage{amsmath,amssymb,amsfonts}
\usepackage{algorithmic}
\usepackage{graphicx}
\usepackage{textcomp}
\usepackage{epsfig} % for postscript graphics files
\usepackage{times} % assumes new font selection scheme installed
\usepackage{nicefrac}
\usepackage{bm}
\usepackage{enumerate}
\usepackage{caption}
\usepackage{subcaption} % This may cause reference issues (comment out -> compile -> put back in)
\usepackage{float}
\usepackage[table,xcdraw,dvipsnames]{xcolor}
\usepackage{tabularx}
\usepackage{cancel}
\usepackage[normalem]{ulem}

\def\BibTeX{{\rm B\kern-.05em{\sc i\kern-.025em b}\kern-.08em T\kern-.1667em\lower.7ex\hbox{E}\kern-.125emX}}

\graphicspath{ {./images/} }
\begin{document}

%%%%%%%%%%%%%%%%%%%%%%%%%%%% TITLE %%%%%%%%%%%%%%%%%%%%%%%%%%%%%%%%%%%%
\title{\LARGE \bf
    % \vspace{1em}
    A Mapping of Assurance Techniques for Learning Enabled Autonomous Systems to the Systems Engineering Lifecycle
}

\author{
    Christian Ellis$^{1}$, Maggie Wigness$^{2}$, and Lance Fiondella$^{1}$
    \thanks{
        $^{1}$
        Christian Ellis is a PhD Student and Lance Fiondella is an Associate Professor in the Department of Electrical and Computer Engineering at the University of Massachusetts Dartmouth, USA.
        {\tt\small cellis3, lfiondella@umassd.edu}
    }
    \thanks{
        $^{2}$
        Maggie Wigness is a researcher at the United States Army Research Laboratory (ARL).
        % {\tt\small b.d.researcher@army.mil}
    }
}
\maketitle

%%%%%%%%%%%%%%%%%%%%%%%%%%%%%%%%%%%%%%%%%%%%%%%%%%%%%%%%%%%%%%%%%%%%%%%%%%%
% IEEE-ICAA 2021: https://iaa.jhu.edu/icaa/scope-topics.html
% 4-6 Pages
%%%%%%%%%%%%%%%%%%%%%%%%% ABSTRACT %%%%%%%%%%%%%%%%%%%%%%%%%%%%%%%%%%%%%%
\begin{abstract}
Learning enabled autonomous systems provide increased capabilities compared to traditional systems.
However, the complexity of and probabilistic nature in the underlying methods enabling such capabilities present challenges for current systems engineering processes for assurance, and test, evaluation, verification, and validation (TEVV).
This paper provides a preliminary attempt to map recently developed technical approaches in the assurance and TEVV of learning enabled autonomous systems (LEAS) literature to a traditional systems engineering v-model.
This mapping categorizes such techniques into three main approaches: development, acquisition, and sustainment.
We review the latest techniques to develop safe, reliable, and resilient learning enabled autonomous systems, without recommending radical and impractical changes to existing systems engineering processes.
By performing this mapping, we seek to assist acquisition professionals by (i) informing comprehensive test and evaluation planning, and (ii) objectively communicating risk to leaders.
\end{abstract}

%%%%%%%%%%%%%%%%%%%%%%%%% INTRODUCTION %%%%%%%%%%%%%%%%%%%%%%%%%%%%%%%%%%
\section{Introduction}
%PARAGRAPH I.1 - What is the large scale problem (scope)?
It is widely recognized~\cite{freeman2020test} that the complexity and resulting capabilities of autonomous systems created using machine learning methods, which we refer to as learning enabled autonomous systems (LEAS), pose new challenges to systems engineering compared to their traditional counterparts.
Moreover, the inability to translate qualitative assessments to quantitative metrics which measure system performance hinder adoption.
Such limitations make it difficult to produce reliable systems, and even harder to assure~\cite{denney2018towards}.
Without understanding the capabilities and limitations of existing assurance techniques, defining safety and performance requirements that are both clear and testable remains out of reach.

%PARAGRAPH I.2 - What is the smaller problem your attacking?
Mature test, evaluation, verification, and validation (TEVV) methods have been in use for decades to ensure the safety analysis and acquisition of hardware systems~\cite{dau2016test}, but fewer TEVV methods for software are available, and even fewer for software that improves itself through learning~\cite{clark2015autonomy}.
Initial approaches to autonomous systems use control theory to physically model the world and its underlying dynamics~\cite{nise2020control}, while LEAS infer and generalize statistical patterns, which lead to the achievement of goals from a sample of pre-collected training data points.
However, due to the nature of the environments where LEAS are fielded and the massive size of their underlying state spaces, systems will likely encounter states during operation they have never experienced before, yet still being required to take action.

% Figure 1
\begin{figure}[!h]
\centerline{
    \includegraphics[scale=0.78]{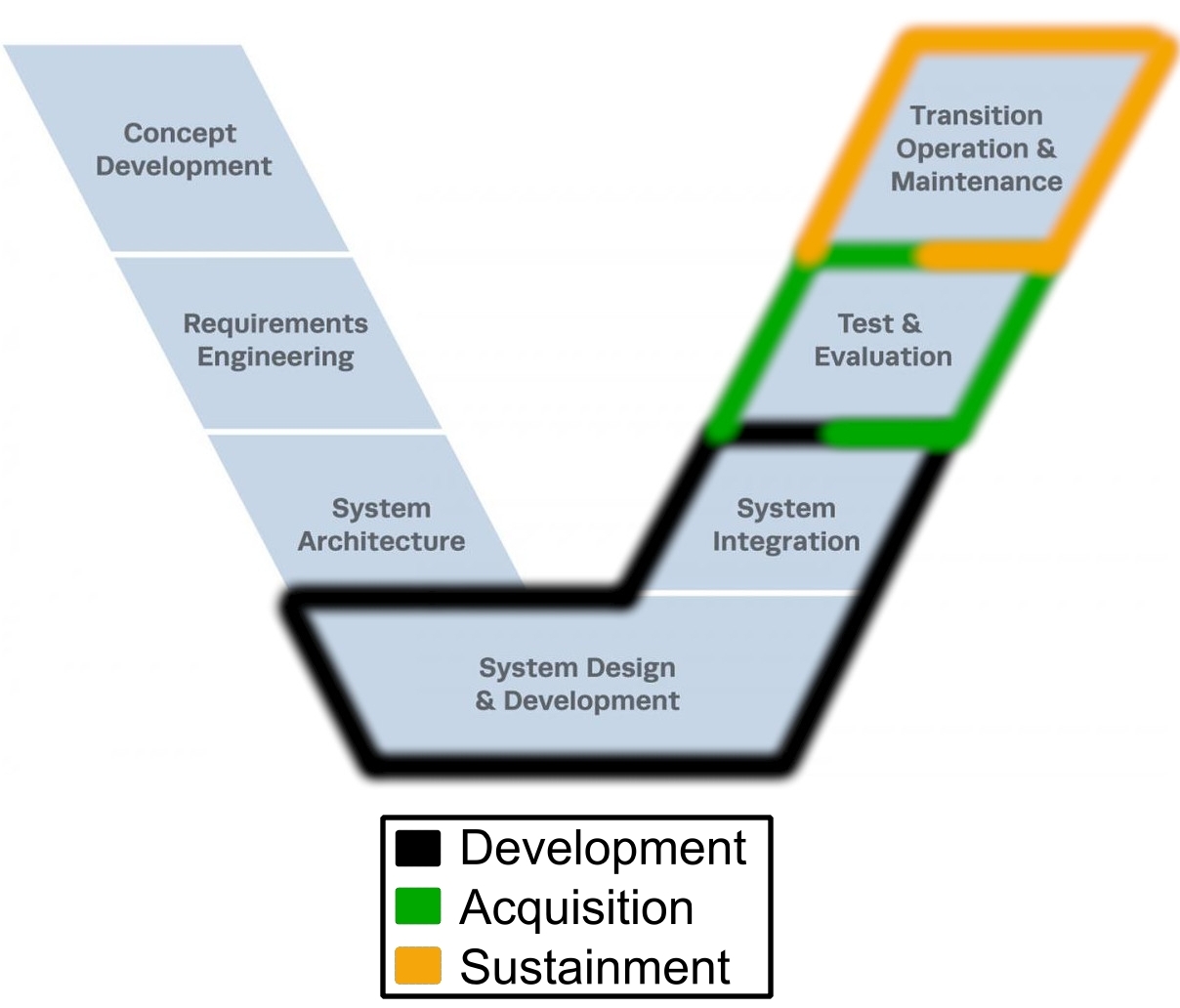}
}
\caption{Recent work in assurance for LEAS is mapped to relevant stages of the MITRE systems engineering lifecycle~\cite{metzger2014systems}, into 3 distinct categories\textemdash development, acquisition, and sustainment.}
\label{fig_v_model}
\end{figure}

% Paragraph I.3 -  How are others addressing the problem and what are the shortcomings and open questions that their solution isn't solving / still missing?
Early work from the autonomy community identified issues that arise from incorporating learning into autonomous systems~\cite{clark2015autonomy} including state space explosion, operation in unpredictable environments, emergent behavior, and effective human machine interaction.
Assurance methods such as formal methods~\cite{baier2008principles}, or reliability analysis~\cite{trivedi2017reliability} seek to provide either certain or probabilistic guarantees on system performance.
Formal methods support verification by exhaustively search and identifying dangerous regions of the state space and provide techniques to avoid such states.
Reliability analysis supports test and evaluation by quantifying the probability a system will be operational at a point in time from operational data collected throughout the systems lifecycle.
The aforementioned state space explosion makes formal methods challenging to scale, and reliability analysis difficult to accurately predict estimates.
A different field of research seeks to develop methods which explicitly consider safety during the learning and operational stages~\cite{amodei2016concrete}.
Lastly, investments such as the DARPA Assured Autonomy program\footnote{https://www.darpa.mil/program/assured-autonomy} seeks to continually assure learning enabled cyber-physical systems by constructing formal methods that assure correctness at design time and perform runtime monitoring at operation time.
While this program has advanced the state of the art in formal methods~\cite{botoeva2020efficient} and runtime monitoring~\cite{beland2020towards}, a systematic approach to identify outstanding gaps will remain unclear unless the community makes explicit and coordinated efforts to understand how such methods may be incorporated into the broader systems engineering process.

%PARAGRAPH I.4 - What are you proposing to address the gap identified?
Our work seeks to communicate recent technical developments in LEAS assurance with a focus on autonomous vehicles, accompanying recent literature reviews~\cite{batarseh2021survey}~\cite{porter2020test}, by mapping such developments to distinct steps of a well known systems engineering model chosen due to its prevalence, namely the v-model.
Fig.~\ref{fig_v_model} shows the mapping and identifies three top level lifecycle phases: development, acquisition, and sustainment. 
For each top level lifecycle phase, a section of the paper has been dedicated to outlining recent technical developments and how they contribute to the goals of the phase.
This representation helps identify where the latest methods for TEVV fit in the broader systems engineering process while also enabling systematic consideration of potential sources of defects, faults, and attacks.
Note that we use the v-model only to assist the classification of where TEVV methods fit. This is not a recommendation to use a certain software development lifecycle over another.

%PARAGRAPH I.5 - Paper Outline
The remainder of the paper is organized as follows. Section~\ref{sec:methods_for_AIAV} outlines the specific scientific fields supporting LEAS.
Section~\ref{sec:framework} provides an overview of the mapping between traditional systems engineering and the state of the art in assurance for LEAS.
Section~\ref{sec:assurance_development} maps assurance techniques, which assist development to \emph{design and development}, and \emph{system integration} in the systems engineering lifecycle.
Section~\ref{sec:assurance_acquisition} maps assurance techniques, which assist acquisition to \emph{test and evaluation}.
Section~\ref{sec:assurance_sustainment} maps assurance techniques, which assist sustainment to \emph{transition, operation, \& maintenance}.
Lastly, section~\ref{sec:conclusion} concludes with areas this mapping can impact.

\section{Methods Supporting the Development of Learning Enabled Autonomous Systems} \label{sec:methods_for_AIAV}
This section seeks to define the fields of engineering with significant impact on development of LEAS with a focus on vehicles and their corresponding challenges for assurance.
In later sections (Sec.~\ref{sec:assurance_development}\textemdash \ref{sec:assurance_sustainment}), solutions to such challenges are identified and categorized according to where they reside within the systems engineering lifecycle such as development, acquisition, or sustainment.
Rather than seek to obtain an exhaustive list of engineering fields, of which there are many, we first provide an overview of learning enabled autonomous vehicles and then review two key contributing fields, including machine learning and reinforcement learning.
While there are other non-learning methods such as optimal control theory~\cite{kirk2004optimal}, which have made large and long lasting impacts on the development on LEAS, they are not considered in this paper.

LEAS normally follow one of two design approaches, end-to-end (E2E) or modular.
In the E2E approach~\cite{tampuu2020survey}, a system's sensors act as the input to a learning algorithm. 
For example, a deep neural network outputs the corresponding actions such as steering wheel angle (lateral control) ~\cite{bojarski2016end}, torque (longitudinal control)~\cite{george2018imitation}, or both~\cite{yu2017baidu}.
In the modular approach~\cite{yurtsever2020survey}, a system's sensors act as input to a perception sub-system which is responsible for building a map and model of the world.
Such subsystems commonly include perception components that use ML techniques such as semantic segmentation~\cite{minaee2021image} or object detection~\cite{zou2019object}.
This model is then used by a planning subsystem, which outputs a kinematically feasible trajectory to which controls are applied~\cite{urmson2008autonomous, guizzogoogle}.

While it may be possible to break down layers of E2E neural networks into sub-components using interpretability techniques~\cite{fan2021interpretability}, this paper specifically focuses on the modular approach for two reasons: i) it is clear to a human what the responsibility of each component is (increased interpretability), and ii) the modular approach is currently more common in autonomous vehicle designs in industry and government implementations.
While some of the underlying problems for assurance are the same for both approaches, including those previously mentioned~\cite{clark2015autonomy}, we explicitly consider software assurance methods which are applicable to either perception or planning components, or the joint-combination thereof.

\subsection{Machine Learning}
In machine learning (ML), tasks are completed by training a model from data to perform function approximation using a combination of mathematical optimization and statistical techniques~\cite{bishop2006pattern}.
This results in computer programs which are able to complete a task without constructing a set of exact solution instructions ahead of time.
There are three main forms of learning, including supervised, unsupervised, and reinforcement learning.
In supervised learning, each training sample from the dataset is associated with a set of features and a corresponding label to train a model.
For example, a neural network can be trained on a dataset of images containing handwritten digits, where each sample's corresponding label is $0$ through $9$.
In unsupervised learning, each training sample is only represented by a set of extracted features, which are subsequently used to identify the underlying feature patterns throughout the dataset.
For example, clustering techniques divide a dataset into $k$ distinct groups, where all data points in a group are similar with respect to some distance measure.
Finally, in reinforcement learning, an autonomous agent learns the optimal way to act over time via interaction with the environment, such as an autonomous robot learning how to move its actuators and joints to navigate in an environment without hitting obstacles.

Although the ability to perform complex tasks solely from data has made ML highly successful, it is for this same reason that ML models are difficult to assure.
Among other factors, a model's performance depends on the data experienced during training and the environment in which it was trained~\cite{aimil}.
Naive metrics such as the model's accuracy on a test set may be perceived as overconfident because they assume most future data will be like the experienced data.
This is especially true in complex systems such as government systems tasked with operating in contested operational environments, demonstrating the need for metrics to assess model performance in new environments.
Another assurance challenge includes determining relevant test cases given the state space explosion and curse of dimensionality problems, of which the Range Adversarial Planning Tool has been proposed~\cite{mullins2017delivering}.
Furthermore, such models are brittle to perturbations in input, which may come from sources such as, sensor noise or adversarial attacks~\cite{brown2017adversarial}.
Lastly, it is inevitable that such models will fail from time to time, and explanations of why they fail (interpretability techniques) and how to fail gracefully (resilience techniques) are also valuable.
Although there are a variety of new assurance techniques~\cite{batarseh2021survey}~\cite{porter2020test} that seek to alleviate such issues, a framework does not exist to assess their thoroughness and relative effectiveness.

\subsection{Reinforcement Learning}
Reinforcement learning (RL) is given a dedicated subsection because it is an enabler of intelligent-like capabilities required for complex autonomous systems.
Reinforcement learning provides a framework for autonomous agents to make decisions under uncertainty and learn from environmental interaction~\cite{sutton2018reinforcement}.
Specified by a reward function, an agent seeks to obtain an optimal policy which maximizes its reward by taking actions over a time horizon in an environment.
A policy is a function that maps the current state to the single action that maximizes the expected future reward.
Formally, this structure is part of a Markov Decision Process (MDP)
consisting of a state space $\mathcal{S}$, an action space $\mathcal{A}$, a state transition distribution over next states $T(s_{t+1}|s_t, a)$, and a reward function $R(s,a,s^\prime)$ whose solution is the optimal policy which maximizes the expected future reward $\pi^*$.
Exact RL seeks to converge to the optimal policy using tabular techniques, requiring an agent to visit each state many times.
Conversely, approximate techniques such as deep RL~\cite{arulkumaran2017brief} allow an agent to operate in large (possibly infinite) state and action spaces without explicitly visiting each state by obtaining a parameterized policy.

Although RL has demonstrated its ability to mimic intelligent capabilities such as beating players at Go~\cite{silver2016mastering}, and autonomous driving~\cite{kiran2021deep}, there are limitations.
Designing reward functions explicitly by hand is a challenging task that can lead to a misalignment between the reward function specified and the true reward function the algorithm designer intended~\cite{russell2019human}.
Such value misalignment leads to unintended consequences such as reward hacking~\cite{clark_amodei_2016}, where the robot maximizes reward in a way that the algorithm designer did not intend while often failing to meet its goals.
Furthermore, many solutions sample inefficiently and are often brittle~\cite{haarnoja2018soft}, limiting their real world applicability.
Lastly, RL can cause a disconnect between how a programmer may interpret what an agent has learned and the true learned concept~\cite{koch2021objective}.
For example, a programmer may believe an agent has learned to traverse to a goal grid cell, but because of the environment setup, the agent may have actually simply learned to traverse to a green grid cell.
For systems incorporating RL, such limitations and corresponding tests for each should be considered explicitly in the assurance process.

%%%%%%%%%%%%%%%%%%%%%%%%%%%%%%%%%%%%%%%%%%%%%%%%%%%%%%%%%%%%%%%%%%%%%%%%%%%
\section{Overview of Mapping}\label{sec:framework}
This paper provides a preliminary attempt to map recently developed technical approaches in the assurance and TEVV of learning enabled autonomous systems (LEAS) literature to a traditional systems engineering v-model.
The mapping identifies three top level lifecycle phases: development, acquisition, and sustainment.
Proceeding according to the colors in Fig. ~\ref{fig_v_model}, the stages surrounded in the black box, including \emph{system design \& development}, and \emph{system integration}, assist development and therefore are mapped to methods which explicitly provide safety assurance during the learning process (Sec.~\ref{sec:assurance_development}).
The stage in the green box, \emph{test and evaluation}, assists the acquisition of systems and therefore is mapped to TEVV analysis techniques which quantify the performance of an already built system, or component (Sec.~\ref{sec:assurance_acquisition}).
The stage in the orange box, \emph{transition operation \& maintenance}, is mapped to safety assurance techniques which aid sustainment by monitoring or adapt performance of a fielded system (Sec.~\ref{sec:assurance_sustainment}).
For each stage, applicable classes of techniques are organized by respective subsections.

In addition to the stage of the system engineering lifecycle, this mapping also seeks to categorize technical developments according to their granularity.
When evaluating different approaches to the same problem, the choice of performance metrics depend on the scope of the unit under test\textemdash whole system, learning enabled component, or a traditional component.
Interfaces at various lifecycle levels of granularity promote systems thinking~\cite{rechtin2010art} about architecture.
Namely, the way a system's components and subsystems relate, interact, and work over time.
By understanding the input paths that contribute to a unit's decisions, the outputs that may lead to failures within the larger system become clearer.

Generally speaking, there are two main approaches to assure LEAS\textemdash white-box techniques and black-box techniques.
White-box techniques require either a model of the system under test, or direct access to the source code.
In contrast, black-box techniques only look at the inputs and outputs of the system under test, and are unaware of the underlying methods of how the system generates the outputs.
White-box techniques are better for component level assurance, while black-box techniques are often better for system-wide assurance.

The implementation of assurance techniques and their accompanying metrics to quantify system performance and safety (Sec.~\ref{sec:assurance_development}\textemdash Sec.~\ref{sec:assurance_sustainment}) can all be used as supporting evidence for a safety assurance case\cite{koopman2019safety} to determine system readiness level and maturity.
Tools which automate traceability and reproducibility throughout the system lifecycle such as~\cite{hartsell2019cps} can reduce the burden of collecting evidence.
The appropriate choice of assurance methods and associated metrics is dependent on the system maturity.
Initial project milestones may focus on demonstrating anti-fragility, while later milestones may focus on demonstrating the ability to accomplish a mission and accompanying capabilities.
Furthermore, quantitative metrics may only be applicable at certain levels of system granularity.
For example, an entire system may be best evaluated by the outcomes of a pre-determined mission and supporting data, while a learning enabled component may be better evaluated by measures specific to machine learning such as uncertainty quantification, robustness to environmental shift, and the ability to fail gracefully and recover from faults.
Metrics which are able to capture the performance of all approaches under test may be preferred over metrics that measure the performance of a certain class of algorithms.
Lastly, if performance data can be collected during the development process, one could also perform a quantitative analysis of a system at any given time using traditional reliability~\cite{trivedi2017reliability} and defect removal~\cite{nafreen2020connecting}.

%%%%%%%%%%%%%%%%%%%%%%%%%%%%%%%%%%%%%%%%%%%%%%%%%%%%%%%%%%%%%%%%%%%%%%%%%%%
\section {Assurance Activities to Support System Development}\label{sec:assurance_development}
This section maps assurance methods which assist development to \emph{system design and development}, and \emph{system integration} in the systems engineering lifecycle.

\subsection{Artificial Intelligence Safety} \label{subsec:ai_saftey}
AI safety is a sub-field of AI which seeks to ensure that a deployed AI systems (i) operates as the designer intended and (ii) completes its task without harming humans.
The importance of AI safety is backed by impactful institutions such as the Future of Life Institute\footnote{https://futureoflife.org/} and Machine Intelligence Research Institute\footnote{https://intelligence.org/}.
In the academic literature, AI safety has been popularized by the agenda of Amodedi et al.~\cite{amodei2016concrete}, who discuss five failure modes for AI; negative side effects, reward hacking, scalable supervision, safe exploration, and distributional shift.
Moreover, in the context of RL, the value alignment problem arises due to a gap in the specified reward function and what the human actually intended~\cite{Hadfield-Menell2021Dissertation}.
Specifically, Taylor et al.~\cite{taylor2016alignment} discuss eight different approaches focusing on two areas of value alignment\textemdash reward specification and techniques to avoid side effects.
Burden et al.~\cite{burden2020exploring} argue that the scope of AI safety problems residing in a specific system can be characterized by three quantitative factors; generality, capability, and control.
For a literature review of AI safety, the reader is directed to~\cite{everitt2018agi}.

While the works above seek to obtain safer agents by altering the underlying methodologies, the focus is on agents in artificial environments rather than physical robots, thereby creating a gap between theoretical and applied research.
Moreover, most approaches assume that the system is following a RL paradigm, demonstrating the importance to understand the underlying learning paradigm employed by a project.
Lastly, although AI safety approaches alone will not be sufficient for LEAS assurance, if the methods are applied during the learning process, such approaches are likely to perform and test better than their non-safe counterparts, leading to higher assurance measures.

\subsection{Learning from Human Feedback}
Incorporating human interaction can positively impact the performance of a LEAS because it is often easier to provide feedback on desired behavior rather than explicitly defining it.
This is one solution to the value alignment problem mentioned in Sec.~\ref{subsec:ai_saftey}.
Such human interaction may include learning from demonstration, intervention, or evaluation~\cite{waytowich2018cycle}.
In learning from demonstration~\cite{ravichandar2020recent}, the human provides a dataset of examples mimicking how the system should operate.
In learning from intervention~\cite{saunders2018trial}, the system operates fully autonomously and the human takes over as required to correct system behavior.
In learning from human evaluation~\cite{christiano2017deep, warnell2018deep}, the system completes various tasks fully autonomously, and then a human ranks the tasks.
This ranking may be from best to worst, or answering the yes/no question, ``Was this the behavior you wanted to see the system perform?'' 
All of the methods mentioned fall into a sub-field known as imitation learning~\cite{osa_2018}.
Lastly, recent developments in imitation learning attempt to incorporate safety as part of the learning process using uncertainty quantification, creating a new sub field known as safe imitation learning~\cite{brown2020safe, ellis2021risk}.

\subsection{Uncertainty Estimation}
System requirements often demand that a learning enabled autonomous system make a prediction, classification, or decision at every time-step during operation.
Since many implementations contain perception systems that will likely never be $100\%$ accurate, the certainty or lack thereof, of a prediction may assist in the final decision made\textemdash especially if the outcome of such a prediction may lead to risky behavior.
The ability for a system to measure what it does and does not know can be captured by quantifying uncertainty with Bayesian analysis techniques~\cite{gelman2013bayesian}.
There are two main types of uncertainty, aleatoric and epistemic.
Aleatoric uncertainty measures the variance between samples in a population.
This type of uncertainty cannot be reduced with more data.
An example is the outcome of a fair coin flip.
Epistemic uncertainty measures the lack of knowledge of a population, which is often captured in a system's parameters.
This type of uncertainty can be alleviated by collecting more data.
An understanding of the different types of uncertainty helps system designers understand if performance can be increased by simply collecting more data.
Additional details can be found in the reviews on uncertainty quantification applied to machine learning~\cite{hullermeier2021aleatoric}, neural networks~\cite{abdar2021review}, and computer vision~\cite{kendall2017uncertainties}.
Such techniques aid at the learning enabled component level, and can be used to quantify system confidence in the current operational environment and thereby communicate uncertainty (risk) to the system end users.

\subsection{Cost-sensitive Learning}
At the system level, the impact of a learning enabled component on the whole system is measured in terms of its ability to assist in the completion of a task.
Additional failure modes introduced by such components must be explicitly considered.
Cost-sensitive learning~\cite{haixiang2017learning} is applicable in classification problems where the cost associated with the misclassification is not equal among classes.
For example, in the context of commercial autonomous vehicles, a false positive resulting in the vehicle stopping when it did not need to likely has lower cost than a false negative resulting in a vehicle colliding with a pedestrian.

\subsection{Formal Methods} \label{subsec:formal_methods}
Static analysis techniques such as formal methods are able to provide guarantees on system performance without ever operating the system~\cite{rival2020introduction}.
Rather than attempting to discover faults while the system is placed under operation, claims about a system are proved or disproved algorithmically using rigorous mathematical methods.
Such methods develop a model of the system being tested, such as a finite-state automaton, and then test that model against a set of specifications defined in a formal language.
There are two main approaches, formal verification~\cite{baier2008principles}, which checks if a given system satisfies a set of specifications, while program synthesis seeks to construct a system from a set of specifications~\cite{gulwani2017program}.
For a literature review of formal methods in the context of autonomous robotics, the reader is directed to~\cite{luckcuck2019formal}.

In the context of LEAS, the system is often complex and safety is critical, thereby making formal methods an attractive solution.
Specifically, synthesis methods provide a ``correct-by-construction" approach~\cite{topcu2019formal}, where capabilities and required operating conditions such as safety constraints are described as specifications and act as input to a synthesis algorithm which outputs the appropriate system model and optimal control policy.
The vehicle's actions are thereby guaranteed to stay within the operating conditions determined by the obtained policy.
However, many approaches are currently limited to static environments, meaning a robot which is guaranteed to satisfy the specifications in one environment does not necessarily carry over to other environments.
Moreover, many formal methods have issues scaling to large state spaces~\cite{fisher2013verifying} due to their exhaustive nature. However, solutions have been proposed using clever optimization techniques such as ~\cite{dehnert2017storm}~\cite{suilen2020robust}.
Nevertheless, synthesis methods can be used to assure safety during a systems development phase, while formal verification techniques such as model checking~\cite{baier2008principles} may be more applicable at the acquirement level.

%%%%%%%%%%%%%%%%%%%%%%%%%%%%%%%%%%%%%%%%%%%%%%%%%%%%%%%%%%%%%%%%%%%%%%%%%%%
\section{Assurance Activities to Support System Acquisition}\label{sec:assurance_acquisition}
This section maps assurance activities to support system acquisition to \emph{test and evaluation} in the systems engineering lifecycle.

\subsection{Autonomy Standards}\label{sec:rw:autonomy_standards}
Standards seek to provide safety assurances, verify capabilities, and promote understanding.
Several standards have been developed to assist the design and development of commercial autonomous vehicles such as ISO 26262~\cite{iso201126262} and IEC 61508\cite{iec61508}.
Specifically, UL 4600~\cite{ul4600} and ISO/PAS 21448~\cite{iso2019pas} explicitly consider autonomous vehicle capabilities incorporating learning.
UL 4600 employs the idea of safety assurance cases, where system performance is argued like a court case given evidence.
Minimizing risk is the goal while also accepting that it cannot be eliminated all together.
Military focused standards include ALFUS~\cite{huang2005framework} paired with the updated ARP6128~\cite{ARP6128}, and MIL-STD-882E~\cite{MIL-STD-882E}.
Most recently, IEEE 2817~\cite{IEEE2817} (in development) seeks to standardize verification methods specifically for autonomous systems.
Although this discussion is part of the development subsection, the standards listed here may also be applicable to the other two lifecycle categories identified in Figure~\ref{fig_v_model}.

\subsection{Software Testing}
Capabilities of autonomous systems are enabled by software.
There is no debate on the importance of software testing, when acknowledging the severity of historic software failures such as the patriot missile or Boeing 737 MAX.
Traditional methods such as those outlined in~\cite{mathur2013foundations} seek to partition the input space using graph or logic coverage to exhaustively test a program.
While traditional methods may work for testing traditional software systems, exhaustive methods are rendered infeasible due to the state space explosion problem.
Moreover, for statistical learning algorithms commonly applied in machine learning methods, the set of all possible samples is often much larger than the number of samples collected.
For example, the set of all possible images a camera may sense using the RGB spectrum with an image size of $256 \times 256$ is ${16,777,216}^{(256*256)}$.
This demonstrates the importance of analyzing dataset features and their associated effectiveness~\cite{barbiero2020modeling} to obtain a generalized model.

In regards to software engineering\textemdash a machine learning model is similar to traditional components, they both have inputs and outputs.
The difference is the size of the input space and that the outputs may change on the same input at different points in time if the model is continually learning.
However, if the model is not learning from new data, it can be considered as a deterministic component.
An inaccurate prediction from a model can be thought of as equivalent to a software fault~\cite{gula2020software}.
However, due to the large state space, the issue remains in the detection of such faults.
The next subsection that follows seek to identify such faults.

\subsection{Automated Test Generation}
Automated test case generation seeks to increase the effectiveness of test and evaluation by minimizing testing time, and identifying the most impactful test cases which are likely to contain faults.
A survey of automatic test-case generation ~\cite{anand2013orchestrated} identifies five main categories\textemdash structural testing, model-based testing, combinatorial testing, random testing, and search-based testing.
Aforementioned for LEAS, the number of configurations is often intractable and therefore exhaustive or tree methods are infeasible.
Search-based methods seek to alleviate this issue by using clever optimization techniques, which identify test cases in areas (boundaries) of a systems configuration space that are likely to lead to system failure.
Therefore, this subsection focuses on search-based methods.

Most relevant to LEAS, Mullins et al.~\cite{mullins2017delivering} provide a tool which automatically identifies test cases for a system under test with a search based optimization approach dependent on a set of mission scenario configurations and a performance score for each configuration.
A case study using the aforementioned tool in an autonomous surface vessel domain can be found in~\cite{stankiewicz2019improving}.
Bridging the gap between formal verification and automated test case generation, Akellea et al.~\cite{akella2020formal} provide a black-box method to identify test cases which do not satisfy a provided temporal logic specification based on a dataset of observed demonstrations.
Most recently, Badithela et al.~\cite{badithela2021synthesis} identify test cases for mission objectives by constructing a set of constraints given a user-defined sequence of waypoints and a reachability objective.
In conclusion, recent research in search-based automated test generation is able to handle the state space explosion problem, while also finding the most impactful test cases.
The results from such test cases help provide impactful evidence towards, or against the construction of safety case (UL 4600).

\subsection{Metrics for Machine Learning}
Metrics provide a quantitative analysis of performance, clearly identifying the best solution out of a set of possible solutions.
Performance is best measured by the system's ability to accurately make predictions in the current operational environment which positively contribute to the larger mission goals.
Initial metrics in supervised machine learning focused on confusion matrices and receiver operating characteristics (ROC) with metrics such as accuracy, sensitivity,  specificity, precision, and F1 score.
For regression models, statistical measures such as mean absolute error or mean squared error were sufficient.
In the context of neural network regression, the statistical significance of input features and an accompanying statistical test may be identified~\cite{horel2020significance}.
In the reinforcement learning framework, Chan et al.~\cite{chan2019measuring} provide a set of test and evaluation metrics to statistically measure the variability and risk of RL algorithms both during and after training.

Agnostic to the task (classification, regression, clustering, etc.), neuron coverage was introduced~\cite{pei2017deepxplore}, as a testing metric analogous to code coverage~\cite{miller1963systematic} for traditional systems.
Code coverage measures the percentage of a code base that has been covered by tests.
High code coverage implies that few software bugs remain, and vice versa.
Similarly, neural coverage measures the percentage neuron activations occur from the testing dataset, seeking to obtain the same implications of code coverage.
However, recent research~\cite{abrecht2020revisiting}~\cite{ harel2020neuron} has shown that neuron coverage is an insufficient metric for testing.
Wang et al.\cite{wang2021robot} seek to address this limitation by quantifying the value of a test set.
Addition metrics research is needed to quantitatively measure the assurance problems outlined in~\cite{clark2015autonomy}.

\section{Assurance Activities to Support System Sustainment} \label{sec:assurance_sustainment}
This section maps assurance activities which support system sustainment to \emph{transition, operating, and maintenance} in the systems engineering lifecycle.

\subsection{Runtime Monitoring}
Runtime monitoring observes the current state of a system and determines if the system is satisfying or violating a set of pre-determined specifications.
This is similar to the formal methods approach outlined in Sec.~\ref{subsec:formal_methods}.
However, runtime monitoring occurs online (while the system is operating), whereas most techniques from formal methods occur offline.
Kane et al. introduced EgMon~\cite{kane2015case}, which detects the violation of specifications using propositional metric temporal logic.
Similarly, Zapridou et al.~\cite{zapridou2020runtime} develop an adaptive cruise control system in the CARLA simulator and perform runtime monitoring using signal temporal logic.
Yel et al.~\cite{yel2020assured} provide a runtime monitoring technique using neural networks for safe motion planning.
In the U.S. government sector, a Boeing team as part of the DARPA assured autonomy program, implemented runtime monitoring in a flight simulator~\cite{beland2020towards}.
For an overview of runtime monitoring techniques, the reader is directed to~\cite{cassar2017survey}.

\subsection{Resilience Engineering}
Resilience engineering techniques seek to build systems which remain operational subject to faults and disturbances~\cite{madni2009towards}.
Such techniques quantify the impact of degraded performance and robustness to faults while providing predictions such as the expected time until recovery~\cite{hosseini2016review}.
In the context of LEAS, resilience techniques can accommodate sensor inaccuracies which may come from measurement limitations in the hardware, dust or debris, and adversarial attacks~\cite{brown2017adversarial}.
Resilience monitoring enables a system to recognize that performance is degraded, and then adapt appropriately, such as moving from perception based navigation to odometry based navigation.
At the time of writing there is little technical research on the incorporation of resilience techniques to LEAS~\cite{johnsen2020review}, However, \cite{prorok2021beyond} offers an initial taxonomy on resilience for multi-robot systems.

%%%%%%%%%%%%%%%%%%%%%%%%%%%%%%%%%%%%%%%%%%%%%%%%%%%%%%%%%%%%%%%%%%%%%
\section{Conclusion} \label{sec:conclusion}
% C.1 - Summary
This paper provides preliminary attempt to map recently developed technical approaches for the assurance of LEAS to a traditional systems engineering v-model.
% C.2 - Importance of work and its impacts
By doing so, we seek to improve the acquisition process by: (i) informing comprehensive assurance planning, (ii) promoting detailed analysis of alternatives, and (iii) objectively communicating risk to leaders.
As indicated by the number of references in each section, most research has been done in the development of methods which explicitly consider safety assurance, while further research is needed in methods which aid the acquirement, and sustainment of such systems.
% C.3 - Future Work
Future work seeks to perform a case study assuring a LEAS using some of the methodologies referenced in this paper.

%%%%%%%%%%%%%%%%%%%%%%%%%%%%%%%%%%%%%%%%%%%%%%%%%%%%%%%%%%%%%%%%%%%%%%%%%%%
% BIBLIOGRAPHY
\bibliographystyle{./bibliography/IEEEtran}
\bibliography{references}

\begin{thebibliography}{10}
\providecommand{\url}[1]{#1}
\csname url@samestyle\endcsname
\providecommand{\newblock}{\relax}
\providecommand{\bibinfo}[2]{#2}
\providecommand{\BIBentrySTDinterwordspacing}{\spaceskip=0pt\relax}
\providecommand{\BIBentryALTinterwordstretchfactor}{4}
\providecommand{\BIBentryALTinterwordspacing}{\spaceskip=\fontdimen2\font plus
\BIBentryALTinterwordstretchfactor\fontdimen3\font minus
  \fontdimen4\font\relax}
\providecommand{\BIBforeignlanguage}[2]{{%
\expandafter\ifx\csname l@#1\endcsname\relax
\typeout{** WARNING: IEEEtran.bst: No hyphenation pattern has been}%
\typeout{** loaded for the language `#1'. Using the pattern for}%
\typeout{** the default language instead.}%
\else
\language=\csname l@#1\endcsname
\fi
#2}}
\providecommand{\BIBdecl}{\relax}
\BIBdecl

\bibitem{freeman2020test}
L.~Freeman, ``Test and evaluation for artificial intelligence,''
  \emph{INSIGHT}, vol.~23, no.~1, pp. 27--30, 2020.

\bibitem{denney2018towards}
E.~Denney, G.~Pai, and M.~Johnson, ``Towards a rigorous basis for specific
  operations risk assessment of uas,'' in \emph{2018 IEEE/AIAA 37th Digital
  Avionics Systems Conference (DASC)}.\hskip 1em plus 0.5em minus 0.4em\relax
  IEEE, 2018, pp. 1--10.

\bibitem{dau2016test}
\emph{Test and Evaluation Management Guide}, 6th~ed.\hskip 1em plus 0.5em minus
  0.4em\relax Defense Acquisition University, 2016.

\bibitem{clark2015autonomy}
M.~Clark, J.~Alley, P.~Deal, J.~DePriest, E.~Hansen, C.~Heitmeyer, R.~Nameth,
  M.~Steinberg, C.~Turner, S.~Young \emph{et~al.}, ``Autonomy community of
  interest (coi) test and evaluation, verification and validation (tevv)
  working group: Technology investment strategy 2015-2018,'' Office of the
  Assistant Secretary of Defense for Research and Engineering, Tech. Rep.,
  2015.

\bibitem{nise2020control}
N.~S. Nise, \emph{Control systems engineering}, 8th~ed.\hskip 1em plus 0.5em
  minus 0.4em\relax John Wiley \& Sons, 2020.

\bibitem{metzger2014systems}
L.~S. Metzger, G.~Rebovich~Jr, R.~A. Cormier, S.~J.~T. Norman, D.~L. Schuh,
  P.~A. Smyton, R.~S. Swarz, and F.~C. Wendt, ``Systems engineering guide:
  Collected wisdom from mitres systems engineering experts,'' MITRE CORP
  BEDFORD MA United States, Tech. Rep., 2014.

\bibitem{baier2008principles}
C.~Baier and J.-P. Katoen, \emph{Principles of model checking}.\hskip 1em plus
  0.5em minus 0.4em\relax MIT press, 2008.

\bibitem{trivedi2017reliability}
K.~S. Trivedi and A.~Bobbio, \emph{Reliability and availability engineering:
  modeling, analysis, and applications}.\hskip 1em plus 0.5em minus 0.4em\relax
  Cambridge University Press, 2017.

\bibitem{amodei2016concrete}
D.~Amodei, C.~Olah, J.~Steinhardt, P.~Christiano, J.~Schulman, and D.~Man{\'e},
  ``Concrete problems in ai safety,'' \emph{arXiv preprint arXiv:1606.06565},
  2016.

\bibitem{botoeva2020efficient}
E.~Botoeva, P.~Kouvaros, J.~Kronqvist, A.~Lomuscio, and R.~Misener, ``Efficient
  verification of relu-based neural networks via dependency analysis,'' in
  \emph{Proceedings of the AAAI Conference on Artificial Intelligence},
  vol.~34, no.~04, 2020, pp. 3291--3299.

\bibitem{beland2020towards}
S.~Beland, I.~Chang, A.~Chen, M.~Moser, J.~Paunicka, D.~Stuart, J.~Vian,
  C.~Westover, and H.~Yu, ``Towards assurance evaluation of autonomous
  systems,'' in \emph{Proceedings of the 39th International Conference on
  Computer-Aided Design}, 2020, pp. 1--6.

\bibitem{batarseh2021survey}
F.~A. Batarseh, L.~Freeman, and C.-H. Huang, ``A survey on artificial
  intelligence assurance,'' \emph{Journal of Big Data}, vol.~8, no.~1, pp.
  1--30, 2021.

\bibitem{porter2020test}
D.~J. Porter and J.~W. Dennis, ``Test \& evaluation of ai-enabled and
  autonomous systems: A literature review,'' 2020.

\bibitem{kirk2004optimal}
D.~E. Kirk, \emph{Optimal control theory: an introduction}.\hskip 1em plus
  0.5em minus 0.4em\relax Courier Corporation, 2004.

\bibitem{tampuu2020survey}
A.~Tampuu, T.~Matiisen, M.~Semikin, D.~Fishman, and N.~Muhammad, ``A survey of
  end-to-end driving: Architectures and training methods,'' \emph{IEEE
  Transactions on Neural Networks and Learning Systems}, 2020.

\bibitem{bojarski2016end}
M.~Bojarski, D.~Del~Testa, D.~Dworakowski, B.~Firner, B.~Flepp, P.~Goyal, L.~D.
  Jackel, M.~Monfort, U.~Muller, J.~Zhang \emph{et~al.}, ``End to end learning
  for self-driving cars,'' \emph{arXiv preprint arXiv:1604.07316}, 2016.

\bibitem{george2018imitation}
L.~George, T.~Buhet, E.~Wirbel, G.~Le-Gall, and X.~Perrotton, ``Imitation
  learning for end to end vehicle longitudinal control with forward camera,''
  \emph{arXiv preprint arXiv:1812.05841}, 2018.

\bibitem{yu2017baidu}
H.~Yu, S.~Yang, W.~Gu, and S.~Zhang, ``Baidu driving dataset and end-to-end
  reactive control model,'' in \emph{2017 IEEE Intelligent Vehicles Symposium
  (IV)}.\hskip 1em plus 0.5em minus 0.4em\relax IEEE, 2017, pp. 341--346.

\bibitem{yurtsever2020survey}
E.~Yurtsever, J.~Lambert, A.~Carballo, and K.~Takeda, ``A survey of autonomous
  driving: Common practices and emerging technologies,'' \emph{IEEE access},
  vol.~8, pp. 58\,443--58\,469, 2020.

\bibitem{minaee2021image}
S.~Minaee, Y.~Y. Boykov, F.~Porikli, A.~J. Plaza, N.~Kehtarnavaz, and
  D.~Terzopoulos, ``Image segmentation using deep learning: A survey,''
  \emph{IEEE Transactions on Pattern Analysis and Machine Intelligence}, 2021.

\bibitem{zou2019object}
Z.~Zou, Z.~Shi, Y.~Guo, and J.~Ye, ``Object detection in 20 years: A survey,''
  \emph{arXiv preprint arXiv:1905.05055}, 2019.

\bibitem{urmson2008autonomous}
C.~Urmson, J.~Anhalt, D.~Bagnell, C.~Baker, R.~Bittner, M.~Clark, J.~Dolan,
  D.~Duggins, T.~Galatali, C.~Geyer \emph{et~al.}, ``Autonomous driving in
  urban environments: Boss and the urban challenge,'' \emph{Journal of Field
  Robotics}, vol.~25, no.~8, pp. 425--466, 2008.

\bibitem{guizzogoogle}
E.~Guizzo, ``How google’s self-driving car works,'' \emph{IEEE Spectrum},
  vol.~18, no.~7, pp. 1132--1141, 2011.

\bibitem{fan2021interpretability}
F.-L. Fan, J.~Xiong, M.~Li, and G.~Wang, ``On interpretability of artificial
  neural networks: A survey,'' \emph{IEEE Transactions on Radiation and Plasma
  Medical Sciences}, 2021.

\bibitem{bishop2006pattern}
C.~M. Bishop, \emph{Pattern recognition and Machine Learning}.\hskip 1em plus
  0.5em minus 0.4em\relax Springer, 2006.

\bibitem{aimil}
``Enabling ai data readiness in the department of defense.''

\bibitem{mullins2017delivering}
G.~E. Mullins, P.~G. Stankiewicz, R.~C. Hawthorne, J.~D. Appler, M.~H. Biggins,
  K.~Chiou, M.~A. Huntley, J.~D. Stewart, and A.~S. Watkins, ``Delivering test
  and evaluation tools for autonomous unmanned vehicles to the fleet,''
  \emph{Johns Hopkins APL technical digest}, vol.~33, no.~4, pp. 279--288,
  2017.

\bibitem{brown2017adversarial}
T.~B. Brown, D.~Man{\'e}, A.~Roy, M.~Abadi, and J.~Gilmer, ``Adversarial
  patch,'' \emph{arXiv preprint arXiv:1712.09665}, 2017.

\bibitem{sutton2018reinforcement}
R.~S. Sutton and A.~G. Barto, \emph{Reinforcement learning: An
  introduction}.\hskip 1em plus 0.5em minus 0.4em\relax MIT press, 2018.

\bibitem{arulkumaran2017brief}
K.~Arulkumaran, M.~P. Deisenroth, M.~Brundage, and A.~A. Bharath, ``Deep
  reinforcement learning: A brief survey,'' \emph{IEEE Signal Processing
  Magazine}, vol.~34, no.~6, pp. 26--38, 2017.

\bibitem{silver2016mastering}
D.~Silver, A.~Huang, C.~J. Maddison, A.~Guez, L.~Sifre, G.~Van Den~Driessche,
  J.~Schrittwieser, I.~Antonoglou, V.~Panneershelvam, M.~Lanctot \emph{et~al.},
  ``Mastering the game of go with deep neural networks and tree search,''
  \emph{nature}, vol. 529, no. 7587, pp. 484--489, 2016.

\bibitem{kiran2021deep}
B.~R. Kiran, I.~Sobh, V.~Talpaert, P.~Mannion, A.~A. Al~Sallab, S.~Yogamani,
  and P.~P{\'e}rez, ``Deep reinforcement learning for autonomous driving: A
  survey,'' \emph{IEEE Transactions on Intelligent Transportation Systems},
  2021.

\bibitem{russell2019human}
S.~Russell, \emph{Human compatible: Artificial intelligence and the problem of
  control}.\hskip 1em plus 0.5em minus 0.4em\relax Penguin, 2019.

\bibitem{clark_amodei_2016}
\BIBentryALTinterwordspacing
J.~Clark and D.~Amodei, ``Faulty reward functions in the wild,'' Dec 2016.
  [Online]. Available: \url{https://openai.com/blog/faulty-reward-functions/}
\BIBentrySTDinterwordspacing

\bibitem{haarnoja2018soft}
T.~Haarnoja, A.~Zhou, P.~Abbeel, and S.~Levine, ``Soft actor-critic: Off-policy
  maximum entropy deep reinforcement learning with a stochastic actor,'' in
  \emph{International conference on machine learning}.\hskip 1em plus 0.5em
  minus 0.4em\relax PMLR, 2018, pp. 1861--1870.

\bibitem{koch2021objective}
J.~Koch, L.~Langosco, J.~Pfau, J.~Le, and L.~Sharkey, ``Objective robustness in
  deep reinforcement learning,'' \emph{arXiv preprint arXiv:2105.14111}, 2021.

\bibitem{rechtin2010art}
E.~Rechtin and M.~W. Maier, \emph{The art of systems architecting}.\hskip 1em
  plus 0.5em minus 0.4em\relax CRC press, 2010.

\bibitem{koopman2019safety}
P.~Koopman, U.~Ferrell, F.~Fratrik, and M.~Wagner, ``A safety standard approach
  for fully autonomous vehicles,'' in \emph{International Conference on
  Computer Safety, Reliability, and Security}.\hskip 1em plus 0.5em minus
  0.4em\relax Springer, 2019, pp. 326--332.

\bibitem{hartsell2019cps}
C.~Hartsell, N.~Mahadevan, S.~Ramakrishna, A.~Dubey, T.~Bapty, T.~Johnson,
  X.~Koutsoukos, J.~Sztipanovits, and G.~Karsai, ``Cps design with
  learning-enabled components: A case study,'' in \emph{Proceedings of the 30th
  International Workshop on Rapid System Prototyping (RSP'19)}, 2019, pp.
  57--63.

\bibitem{nafreen2020connecting}
M.~Nafreen, M.~Luperon, L.~Fiondella, V.~Nagaraju, Y.~Shi, and T.~Wandji,
  ``Connecting software reliability growth models to software defect
  tracking,'' in \emph{2020 IEEE 31st International Symposium on Software
  Reliability Engineering (ISSRE)}.\hskip 1em plus 0.5em minus 0.4em\relax
  IEEE, 2020, pp. 138--147.

\bibitem{Hadfield-Menell2021Dissertation}
D.~Hadfield-Menell, ``The principal-agent alignment problem in artificial
  intelligence,'' Ph.D. dissertation, University of California, Berkeley, 2021.

\bibitem{taylor2016alignment}
J.~Taylor, E.~Yudkowsky, P.~LaVictoire, and A.~Critch, ``Alignment for advanced
  machine learning systems,'' \emph{Ethics of Artificial Intelligence}, pp.
  342--382, 2016.

\bibitem{burden2020exploring}
J.~Burden and J.~Hern{\'a}ndez-Orallo, ``Exploring ai safety in degrees:
  Generality, capability and control,'' in \emph{SafeAI@ AAAI}, 2020.

\bibitem{everitt2018agi}
T.~Everitt, G.~Lea, and M.~Hutter, ``Agi safety literature review,'' in
  \emph{IJCAI}, 2018.

\bibitem{waytowich2018cycle}
N.~R. Waytowich, V.~G. Goecks, and V.~J. Lawhern, ``Cycle-of-learning for
  autonomous systems from human interaction,'' in \emph{AI-HRI Symposium, AAAI
  Fall Symposium Series}, 2018.

\bibitem{ravichandar2020recent}
H.~Ravichandar, A.~S. Polydoros, S.~Chernova, and A.~Billard, ``Recent advances
  in robot learning from demonstration,'' \emph{Annual Review of Control,
  Robotics, and Autonomous Systems}, vol.~3, pp. 297--330, 2020.

\bibitem{saunders2018trial}
W.~Saunders, G.~Sastry, A.~Stuhlm{\"u}ller, and O.~Evans, ``Trial without
  error: Towards safe reinforcement learning via human intervention,'' in
  \emph{Proceedings of the 17th International Conference on Autonomous Agents
  and MultiAgent Systems}, 2018, pp. 2067--2069.

\bibitem{christiano2017deep}
P.~F. Christiano, J.~Leike, T.~B. Brown, M.~Martic, S.~Legg, and D.~Amodei,
  ``Deep reinforcement learning from human preferences,'' in \emph{NIPS}, 2017.

\bibitem{warnell2018deep}
G.~Warnell, N.~Waytowich, V.~Lawhern, and P.~Stone, ``Deep tamer: Interactive
  agent shaping in high-dimensional state spaces,'' in \emph{Thirty-Second AAAI
  Conference on Artificial Intelligence}, 2018.

\bibitem{osa_2018}
T.~Osa, J.~Pajarinen, G.~Neumann, J.~Bagnell, P.~Abbeel, and J.~Peters, ``An
  algorithmic perspective on imitation learning,'' \emph{Foundations and Trends
  in Robotics}, vol.~7, no. 1-2, pp. 1--179, 2018.

\bibitem{brown2020safe}
D.~Brown, R.~Coleman, R.~Srinivasan, and S.~Niekum, ``Safe imitation learning
  via fast bayesian reward inference from preferences,'' in \emph{International
  Conference on Machine Learning}.\hskip 1em plus 0.5em minus 0.4em\relax PMLR,
  2020, pp. 1165--1177.

\bibitem{ellis2021risk}
C.~Ellis, M.~Wigness, J.~G. Rogers~III, C.~Lennon, and L.~Fiondella, ``Risk
  averse bayesian reward learning for autonomous navigation from human
  demonstration,'' in \emph{International Conference on Intelligent Robots and
  Systems, Prague Czech Republic}, 2021.

\bibitem{gelman2013bayesian}
A.~Gelman, J.~B. Carlin, H.~S. Stern, and D.~B. Rubin, \emph{Bayesian data
  analysis}, 3rd~ed.\hskip 1em plus 0.5em minus 0.4em\relax Chapman and
  Hall/CRC, 2013.

\bibitem{hullermeier2021aleatoric}
E.~H{\"u}llermeier and W.~Waegeman, ``Aleatoric and epistemic uncertainty in
  machine learning: An introduction to concepts and methods,'' \emph{Machine
  Learning}, vol. 110, no.~3, pp. 457--506, 2021.

\bibitem{abdar2021review}
M.~Abdar, F.~Pourpanah, S.~Hussain, D.~Rezazadegan, L.~Liu, M.~Ghavamzadeh,
  P.~Fieguth, X.~Cao, A.~Khosravi, U.~R. Acharya \emph{et~al.}, ``A review of
  uncertainty quantification in deep learning: Techniques, applications and
  challenges,'' \emph{Information Fusion}, 2021.

\bibitem{kendall2017uncertainties}
A.~Kendall and Y.~Gal, ``What uncertainties do we need in bayesian deep
  learning for computer vision?'' in \emph{NIPS}, 2017.

\bibitem{haixiang2017learning}
G.~Haixiang, L.~Yijing, J.~Shang, G.~Mingyun, H.~Yuanyue, and G.~Bing,
  ``Learning from class-imbalanced data: Review of methods and applications,''
  \emph{Expert Systems with Applications}, vol.~73, pp. 220--239, 2017.

\bibitem{rival2020introduction}
X.~Rival and K.~Yi, \emph{Introduction to static analysis: an abstract
  interpretation perspective}.\hskip 1em plus 0.5em minus 0.4em\relax Mit
  Press, 2020.

\bibitem{gulwani2017program}
S.~Gulwani, O.~Polozov, R.~Singh \emph{et~al.}, ``Program synthesis,''
  \emph{Foundations and Trends{\textregistered} in Programming Languages},
  vol.~4, no. 1-2, pp. 1--119, 2017.

\bibitem{luckcuck2019formal}
M.~Luckcuck, M.~Farrell, L.~A. Dennis, C.~Dixon, and M.~Fisher, ``Formal
  specification and verification of autonomous robotic systems: A survey,''
  \emph{ACM Computing Surveys (CSUR)}, vol.~52, no.~5, pp. 1--41, 2019.

\bibitem{topcu2019formal}
U.~Topcu, ``Formal specification and correct-by-construction synthesis of
  control protocols for adaptable, human-embedded autonomous systems,''
  Trustees of The University of Pennsylvania Philadelphia United States, Tech.
  Rep., 2019.

\bibitem{fisher2013verifying}
M.~Fisher, L.~Dennis, and M.~Webster, ``Verifying autonomous systems,''
  \emph{Communications of the ACM}, vol.~56, no.~9, pp. 84--93, 2013.

\bibitem{dehnert2017storm}
C.~Dehnert, S.~Junges, J.-P. Katoen, and M.~Volk, ``A storm is coming: A modern
  probabilistic model checker,'' in \emph{International Conference on Computer
  Aided Verification}.\hskip 1em plus 0.5em minus 0.4em\relax Springer, 2017,
  pp. 592--600.

\bibitem{suilen2020robust}
M.~Suilen, N.~Jansen, M.~Cubuktepe, and U.~Topcu, ``Robust policy synthesis for
  uncertain pomdps via convex optimization.''\hskip 1em plus 0.5em minus
  0.4em\relax International Joint Conferences on Artificial Intelligence, 2020.

\bibitem{iso201126262}
I.~ISO, ``26262: Road vehicles-functional safety,'' \emph{International
  Organization for Standardization/FDIS}, vol. 26262, 2011.

\bibitem{iec61508}
I.~IEC, ``61508: Functional safety of electrical/electronic/programmable
  electronic safety-related system,'' \emph{International Electrotechnical
  Commission}, 2010.

\bibitem{ul4600}
U.~UL, ``4600: Standard for safety for the evaluation of autonomous vehicles
  and other products,'' \emph{Underwriters Laboratories}, vol. 4600, 2019.

\bibitem{iso2019pas}
I.~ISO, ``21448: road vehicles-safety of the intended functionality,''
  \emph{International Organization for Standardization}, 2019.

\bibitem{huang2005framework}
H.-M. Huang, K.~Pavek, B.~Novak, J.~Albus, and E.~Messin, ``A framework for
  autonomy levels for unmanned systems (alfus),'' \emph{Proceedings of the
  AUVSI’s unmanned systems North America}, pp. 849--863, 2005.

\bibitem{ARP6128}
SAE, ``Unmanned systems terminology based on the alfus framework,'' vol.
  ARP6128, 2019.

\bibitem{MIL-STD-882E}
U.~D. of~Defense, ``System saftey,'' vol. MIL-STD-882E, 2012.

\bibitem{IEEE2817}
I.~S. Association, ``Guide for verification of autonomous systems,'' vol.
  P2817, 2019.

\bibitem{mathur2013foundations}
A.~P. Mathur, \emph{Foundations of software testing, 2/e}.\hskip 1em plus 0.5em
  minus 0.4em\relax Pearson Education India, 2013.

\bibitem{barbiero2020modeling}
P.~Barbiero, G.~Squillero, and A.~Tonda, ``Modeling generalization in machine
  learning: A methodological and computational study,'' \emph{arXiv preprint
  arXiv:2006.15680}, 2020.

\bibitem{gula2020software}
A.~Gula, C.~Ellis, S.~Bhattacharya, and L.~Fiondella, ``Software and system
  reliability engineering for autonomous systems incorporating machine
  learning,'' in \emph{2020 Annual Reliability and Maintainability Symposium
  (RAMS)}.\hskip 1em plus 0.5em minus 0.4em\relax IEEE, 2020, pp. 1--6.

\bibitem{anand2013orchestrated}
S.~Anand, E.~K. Burke, T.~Y. Chen, J.~Clark, M.~B. Cohen, W.~Grieskamp,
  M.~Harman, M.~J. Harrold, P.~McMinn, A.~Bertolino \emph{et~al.}, ``An
  orchestrated survey of methodologies for automated software test case
  generation,'' \emph{Journal of Systems and Software}, vol.~86, no.~8, pp.
  1978--2001, 2013.

\bibitem{stankiewicz2019improving}
P.~G. Stankiewicz and G.~E. Mullins, ``Improving evaluation methodology for
  autonomous surface vessel colregs compliance,'' in \emph{OCEANS
  2019-Marseille}.\hskip 1em plus 0.5em minus 0.4em\relax IEEE, 2019, pp. 1--7.

\bibitem{akella2020formal}
P.~Akella, M.~Ahmadi, R.~M. Murray, and A.~D. Ames, ``Formal test synthesis for
  safety-critical autonomous systems based on control barrier functions,'' in
  \emph{2020 59th IEEE Conference on Decision and Control (CDC)}.\hskip 1em
  plus 0.5em minus 0.4em\relax IEEE, 2020, pp. 790--795.

\bibitem{badithela2021synthesis}
A.~Badithela and R.~M. Murray, ``Synthesis of static test environments for
  observing sequence-like behaviors in autonomous systems,'' \emph{arXiv
  preprint arXiv:2108.05911}, 2021.

\bibitem{horel2020significance}
E.~Horel and K.~Giesecke, ``Significance tests for neural networks,''
  \emph{Journal of Machine Learning Research}, vol.~21, no. 227, pp. 1--29,
  2020.

\bibitem{chan2019measuring}
S.~C. Chan, S.~Fishman, J.~Canny, A.~Korattikara, and S.~Guadarrama,
  ``Measuring the reliability of reinforcement learning algorithms,'' in
  \emph{International Conference on Learning Representations, Addis Ababa,
  Ethiopia}, 2020.

\bibitem{pei2017deepxplore}
K.~Pei, Y.~Cao, J.~Yang, and S.~Jana, ``Deepxplore: Automated whitebox testing
  of deep learning systems,'' in \emph{proceedings of the 26th Symposium on
  Operating Systems Principles}, 2017, pp. 1--18.

\bibitem{miller1963systematic}
J.~C. Miller and C.~J. Maloney, ``Systematic mistake analysis of digital
  computer programs,'' \emph{Communications of the ACM}, vol.~6, no.~2, pp.
  58--63, 1963.

\bibitem{abrecht2020revisiting}
S.~Abrecht, M.~Akila, S.~S. Gannamaneni, K.~Groh, C.~Heinzemann, S.~Houben, and
  M.~Woehrle, ``Revisiting neuron coverage and its application to test
  generation,'' in \emph{International Conference on Computer Safety,
  Reliability, and Security}.\hskip 1em plus 0.5em minus 0.4em\relax Springer,
  2020, pp. 289--301.

\bibitem{harel2020neuron}
F.~Harel-Canada, L.~Wang, M.~A. Gulzar, Q.~Gu, and M.~Kim, ``Is neuron coverage
  a meaningful measure for testing deep neural networks?'' in \emph{Proceedings
  of the 28th ACM Joint Meeting on European Software Engineering Conference and
  Symposium on the Foundations of Software Engineering}, 2020, pp. 851--862.

\bibitem{wang2021robot}
J.~Wang, J.~Chen, Y.~Sun, X.~Ma, D.~Wang, J.~Sun, and P.~Cheng, ``Robot:
  Robustness-oriented testing for deep learning systems,'' in \emph{2021
  IEEE/ACM 43rd International Conference on Software Engineering (ICSE)}.\hskip
  1em plus 0.5em minus 0.4em\relax IEEE, 2021, pp. 300--311.

\bibitem{kane2015case}
A.~Kane, O.~Chowdhury, A.~Datta, and P.~Koopman, ``A case study on runtime
  monitoring of an autonomous research vehicle (arv) system,'' in \emph{Runtime
  Verification}.\hskip 1em plus 0.5em minus 0.4em\relax Springer, 2015, pp.
  102--117.

\bibitem{zapridou2020runtime}
E.~Zapridou, E.~Bartocci, and P.~Katsaros, ``Runtime verification of autonomous
  driving systems in carla,'' in \emph{International Conference on Runtime
  Verification}.\hskip 1em plus 0.5em minus 0.4em\relax Springer, 2020, pp.
  172--183.

\bibitem{yel2020assured}
E.~Yel, T.~J. Carpenter, C.~Di~Franco, R.~Ivanov, Y.~Kantaros, I.~Lee,
  J.~Weimer, and N.~Bezzo, ``Assured runtime monitoring and planning: toward
  verification of neural networks for safe autonomous operations,'' \emph{IEEE
  Robotics \& Automation Magazine}, vol.~27, no.~2, pp. 102--116, 2020.

\bibitem{cassar2017survey}
I.~Cassar, A.~Francalanza, L.~Aceto, and A.~Ing{\'o}lfsd{\'o}ttir, ``A survey
  of runtime monitoring instrumentation techniques,'' \emph{arXiv preprint
  arXiv:1708.07229}, 2017.

\bibitem{madni2009towards}
A.~M. Madni and S.~Jackson, ``Towards a conceptual framework for resilience
  engineering,'' \emph{IEEE Systems Journal}, vol.~3, no.~2, pp. 181--191,
  2009.

\bibitem{hosseini2016review}
S.~Hosseini, K.~Barker, and J.~E. Ramirez-Marquez, ``A review of definitions
  and measures of system resilience,'' \emph{Reliability Engineering \& System
  Safety}, vol. 145, pp. 47--61, 2016.

\bibitem{johnsen2020review}
S.~Johnsen and S.~Kilskar, ``A review of resilience in autonomous transport to
  improve safety and security.''\hskip 1em plus 0.5em minus 0.4em\relax ESREL,
  2020.

\bibitem{prorok2021beyond}
A.~Prorok, M.~Malencia, L.~Carlone, G.~S. Sukhatme, B.~M. Sadler, and V.~Kumar,
  ``Beyond robustness: A taxonomy of approaches towards resilient multi-robot
  systems,'' \emph{arXiv preprint arXiv:2109.12343}, 2021.

\end{thebibliography}
%%%%%%%%%%%%%%%%%%%%%%%%%%%%%%%%%%%%%%%%%%%%%%%%%%%%%%%%%%%%%%%%%%%%%%%%%%%
\clearpage
\end{document}